\documentclass[aaspp4,apjl,apjfonts,natbib209,twocolappendix]{emulateapj}
\pdfoutput=1

\usepackage{graphicx}
\usepackage{times}
 \usepackage{xspace}
\usepackage{amsmath}
\usepackage{mathtools}
\usepackage{stix}

\usepackage[usenames,dvipsnames,svgnames,table]{xcolor}

\usepackage[colorlinks=True, 
	    linkcolor=blue,
            urlcolor=blue,
            citecolor=blue]{hyperref}

\newcommand{\unit}[1]{\mathrm{\,#1}}
\newcommand{\Msun}{\ensuremath{M_{\odot}}\xspace}
\newcommand{\Mstar}{\ensuremath{M_*}\xspace}
\newcommand{\HST}{\textit{HST}}

\shorttitle{Testing The Recovery of Intrinsic Galaxy Sizes and Masses Using Simulations}
\shortauthors{Price et al.}
\slugcomment{Published in ApJL}

\begin{document}

\title{Testing the Recovery of Intrinsic Galaxy Sizes and Masses 
of \lowercase{$z\sim2$} Massive Galaxies\\Using Cosmological Simulations}

\author{Sedona H. Price\altaffilmark{1,*}}
\author{Mariska Kriek\altaffilmark{1}} 
\author{Robert Feldmann\altaffilmark{1,2}}
\author{Eliot Quataert\altaffilmark{1}}
\author{Philip F. Hopkins\altaffilmark{3}}
\author{\\Claude-Andr\'{e} Faucher-Gigu\`{e}re\altaffilmark{4}}
\author{Du\v{s}an Kere{\v{s}}\altaffilmark{5}}
\author{Guillermo Barro\altaffilmark{1}}

\altaffiltext{1}{Department of Astronomy, University of California,
  Berkeley, CA 94720, USA}
\altaffiltext{2}{Institute for Computational Science, University of Zurich, CH-8057 Zurich, Switzerland}
\altaffiltext{3}{TAPIR 350-17, California Institute of Technology, Pasadena, CA 91125, USA}
\altaffiltext{4}{Department of Physics and Astronomy and CIERA, Northwestern University, Evanston, IL 60208, USA}
\altaffiltext{5}{Center for Astrophysics and Space Sciences, University of California, San Diego, CA 92093, USA}
\altaffiltext{*}{email: sedona@berkeley.edu}


\begin{abstract}
Accurate measurements of galaxy masses and sizes are key to tracing galaxy evolution over time.
Cosmological zoom-in simulations provide an ideal test bed for assessing the recovery of 
galaxy properties from observations. 
Here, we utilize galaxies with $M_*\sim10^{10}-10^{11.5}\Msun$ at $z\sim1.7-2$
from the MassiveFIRE cosmological simulation suite, 
part of the Feedback in Realistic Environments (FIRE) project. 
Using mock multi-band images, we compare intrinsic galaxy masses and sizes to observational estimates. 
We find that observations accurately recover stellar masses, with a slight average underestimate of 
 $\sim\!0.06\unit{dex}$ and a $\sim\!0.15\unit{dex}$ scatter.
Recovered half-light radii agree well with intrinsic half-mass radii when 
averaged over all viewing angles, 
with a systematic offset of $\sim\!0.1\unit{dex}$ 
(with the half-light radii being larger) 
and a scatter of $\sim\!0.2\unit{dex}$. 
When using color gradients to account for mass-to-light variations,
recovered half-mass radii also exceed the intrinsic half-mass radii by $\sim\!0.1\unit{dex}$. 
However, if not properly accounted for, aperture effects can bias size estimates by $\sim\!0.1\unit{dex}$. 
No differences are found between the mass and size offsets for star-forming and quiescent galaxies. 
Variations in viewing angle are responsible for $\sim25\%$ of the scatter in the recovered masses and sizes. 
Our results thus suggest that the intrinsic scatter in the mass-size relation may have previously 
been overestimated by $\sim\!25\%$. 
Moreover, orientation-driven scatter causes the number density of 
very massive galaxies to be overestimated by $\sim\!0.5\unit{dex}$ 
at $M_*\sim10^{11.5}\Msun$. 
\end{abstract}

\maketitle\section{Introduction}\setcounter{footnote}{0}


Tracing the evolution of galaxy stellar masses and sizes across multiple cosmological epochs 
provides direct constraints on the growth of galaxies. 
Recent photometric studies have probed stellar masses for 
large galaxy samples out to $z\sim3$ (e.g., \citealt{Tomczak14}), 
and for small samples out to $z\sim9-11$ (e.g., \citealt{Oesch13}). 
Deep, high-resolution \textit{Hubble Space Telescope} (\HST) imaging has also provided 
measurements of rest-frame optical sizes for large samples of galaxies out to 
$z\sim2.5$ (e.g., \citealt{vanderWel14a}, \citealt{Peth16}). 
Together, these measurements make it possible to trace the 
evolution of the mass-size relation \citep{Shen03} out to $z\sim2.5$ \citep{vanderWel14a}.

Despite their central role in galaxy evolution studies, 
it is uncertain how well measured masses and sizes reflect the intrinsic properties of galaxies.
Recovered galaxy properties may be impacted by complex dust-to-star geometry and projection effects. 
Furthermore, galaxy sizes are often measured from the stellar light 
distribution, even though light does not directly trace stellar mass in most galaxies. 
Half-light radii are larger than half-mass radii for many galaxies \citep{Wuyts12}. 
Color gradients can be used to estimate half-mass radii (e.g., \citealt{Szomoru13}), 
but it is unclear how accurately they reflect the intrinsic galaxy sizes.

Evaluating parameter recovery requires a galaxy sample  with known intrinsic properties. 
Mock observations of simulated galaxies are ideally suited to this task, 
as cosmological simulations now probe the complex star, gas, and 
dust geometry in the interstellar medium with high (sub-kiloparsec scale) resolution 
(e.g., \citealt{Hopkins14}, \citealt{Schaye14}, \citealt{Vogelsberger14}, \citealt{Feldmann16}). 
Recent studies have investigated the recovery of stellar masses (e.g., \citealt{Wuyts09b}, 
 \citealt{Hayward15}, \citealt{Torrey15}) and sizes (e.g., \citealt{Wuyts10}, \citealt{Snyder15a,Snyder15}, 
\citealt{Taghizadeh-Popp15}, \citealt{Bottrell17}) using mock observations. 
However, these studies have not simultaneously included dust, multiple viewing angles, 
high spatial resolution, observational point-spread functions (PSFs), and noise 
to test parameter recovery in high-redshift galaxies.

In this Letter, we present a study of the recovery of galaxy masses and sizes using mock observations 
over multiple projections of $z\sim2$ galaxies from MassiveFIRE \citep{Feldmann16}, 
following the same procedures used for observations. 
Throughout this work, we adopt a $\Lambda$CDM cosmology with 
$\Omega_m=0.3$, $\Omega_{\Lambda}=0.7$, and $H_0=70\unit{km}\unit{s^{-1}}\unit{Mpc^{-1}}$.



\begin{figure*}\centering\includegraphics[width=0.95\textwidth]{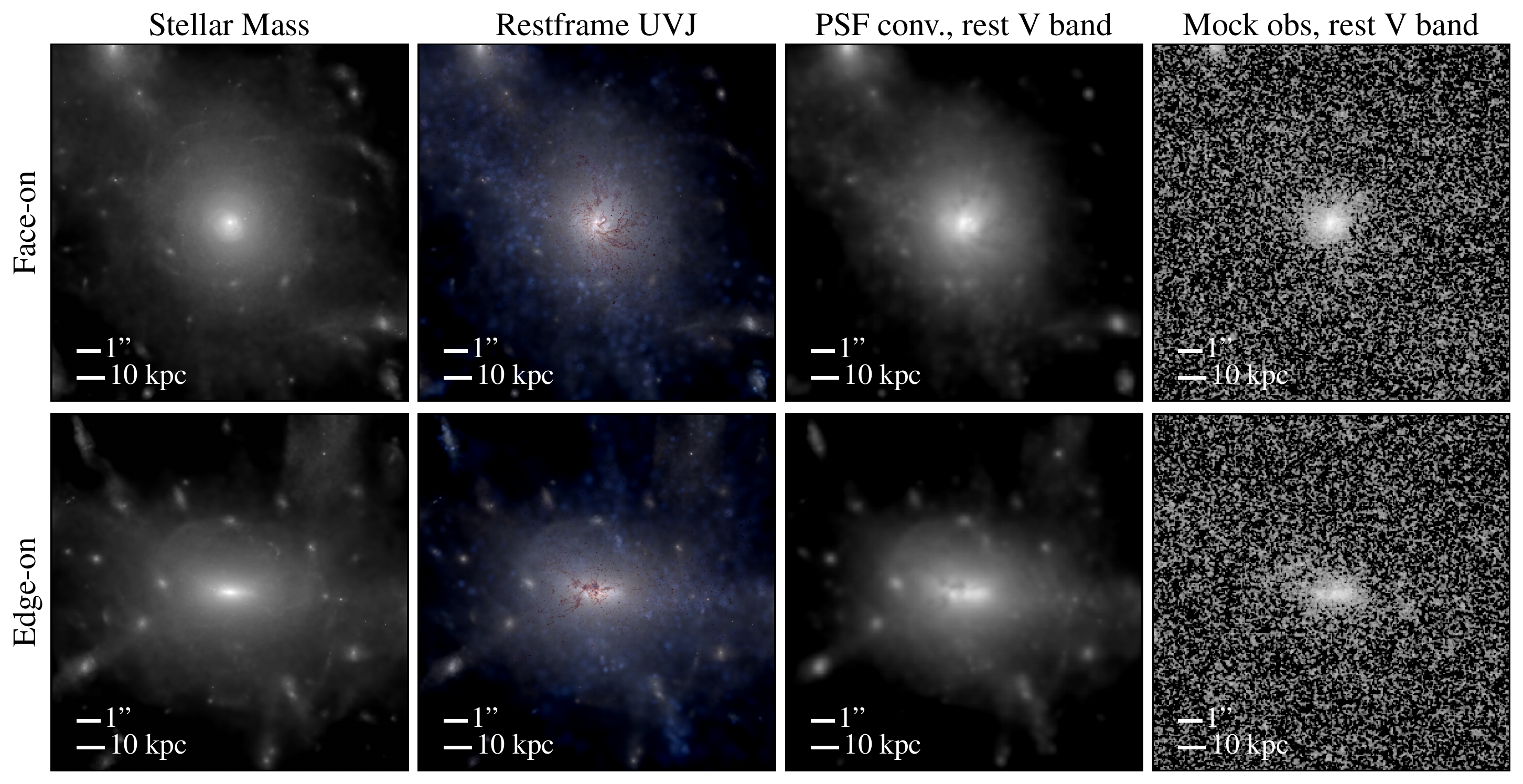}\caption{Example simulated 
		star-forming disk galaxy ($M_*\sim10^{10.9}\Msun,\unit{}r_{1/2,*}\sim4\unit{kpc}$)
		viewed face-on (top) and edge-on (bottom). 
		The first column shows the stellar mass maps. 
		Rest-frame UVJ images (second column) highlight the distribution of dust and stars. 
		We also show the PSF-convolved simulation image (third column) 
		and the resulting mock observation (fourth column) for the rest-frame V band.
		Each image is 144 kpc on each side.}\label{fig:example_mock}\end{figure*}

\section{Mock Observations}
\label{sec:mock_obs}

We use simulations from the FIRE project \citep{Hopkins14} to constrain how well 
intrinsic galaxy properties can be recovered from observations. Specifically, we analyze 
the MassiveFIRE suite of cosmological zoom-in simulations \citep{Feldmann16,Feldmann17}, 
focusing on a sample of 50 massive 
galaxy snapshots. We consider all massive ($M_{*}\sim10^{10}-10^{11.5}\Msun$) 
central and satellite galaxies from the high-resolution runs, 
using snapshots at both $z\sim1.7$ and $z\sim2$ of 
21 galaxies (Series A and B in \citealt{Feldmann17}) and at $z\sim2$ of 8 galaxies 
(Series C in \citealt{Feldmann17}, including 4 unpublished galaxies). 
The sample includes large star-forming disks, irregular star-forming galaxies, and 
quiescent galaxies. 

We construct mock multi-filter images of the galaxies using 
the method described below. To understand how viewing angle affects 
measurements, we generate images of each galaxy along 25 different projections.
First, we generate noise-free multi-filter rest-frame images for each  
projection of each simulated galaxy. 
Every stellar particle is assigned the spectral energy distribution (SED) 
of a simple stellar population based on its mass, age, and metallicity 
using the \citet{bc03} stellar population synthesis (SPS) models assuming a \citet{Chabrier03} 
initial mass function (IMF). 
Dust attenuation is incorporated by tracing the amount of dust 
along the line of sight, assuming a \citet{Calzetti00} curve. 
Dust content is inferred from the gas particle masses and metallicities, 
assuming a fixed dust-to-metal ratio.
Scattering is indirectly applied by using an empirical dust attenuation curve. 
Dust emission is omitted as we do not sample the SEDs at long wavelengths. 
We then sample the dust-attenuated SED in a set of rest-frame filters to obtain mock rest-frame images. 

The images are artificially redshifted to 
the snapshot redshift ($z=2.02$ and $1.67$)
by applying cosmological dimming, adjusting the angular size,
and resampling to match the typical \HST/WFC3 drizzled pixel scale ($0.\arcsec06$). 
The images are convolved with a typical WFC3 PSF (measured from a stack of stars from 
CANDELS \HST/F160W imaging; \citealt{Skelton14}).\footnote{Observationally,  
images are first convolved with the PSF and subsequently sampled within pixels. 
We find no difference in the mock images when 
inverting the calculation order.} 
For simplicity, we apply the same PSF to all bands. 
Mismatches between the PSFs of different photometric bands 
can introduce uncertainties in the relative flux calibration. 
Investigating this uncertainty is beyond the scope of this Letter. 
However, we note that other studies have investigated the 
accuracy of flux recovery from low-resolution photometry 
(e.g., \citealt{Labbe06}, \citealt{Laidler07}, \citealt{Wuyts08}). 
Finally, we add noise in each band using random CANDELS \HST/F160W 
postage stamps, which contain no detected objects in the 
3D-HST catalogs \citep{Skelton14} and have typical noise levels. 
Mock images of each galaxy are constructed for  
16 rest-frame filters: ST-UV14, ST-UV17, ST-UV22, ST-UV27 (from \citealt{bc03}), 
SDSS ugriz,  U, B, V, R, J, H, and K.  
Figure~\ref{fig:example_mock} shows an example face-on and 
edge-on view of one galaxy, demonstrating 
the underlying mass distribution, the rest-frame UVJ colors, 
the PSF-convolved rest-frame V-band image, and the final mock image including noise.

We detect objects and extract photometry from the 
mock images following the procedure by \citet{Skelton14}. 
For every projection of each simulation, 
we use Source Extractor \citep{Bertin96} in dual-image mode, 
adopting the parameters used by \citeauthor{Skelton14} and using the rest-frame V-band for detection 
(roughly covered by F160W at $z\sim2$).  
The multi-band aperture and total photometry and errors of the objects are determined following 
\citeauthor{Skelton14} 
In some projections, dust lanes or bright star-forming clumps lead to multiple detected 
objects for a single galaxy. To account for this issue, we classify all objects with 
segmentation maps falling within $2.5\unit{kpc}$ of the galaxy center as part of the galaxy.

\begin{figure*}\centering\includegraphics[width=0.99\textwidth,trim=0 6pt 0 0,clip=True]{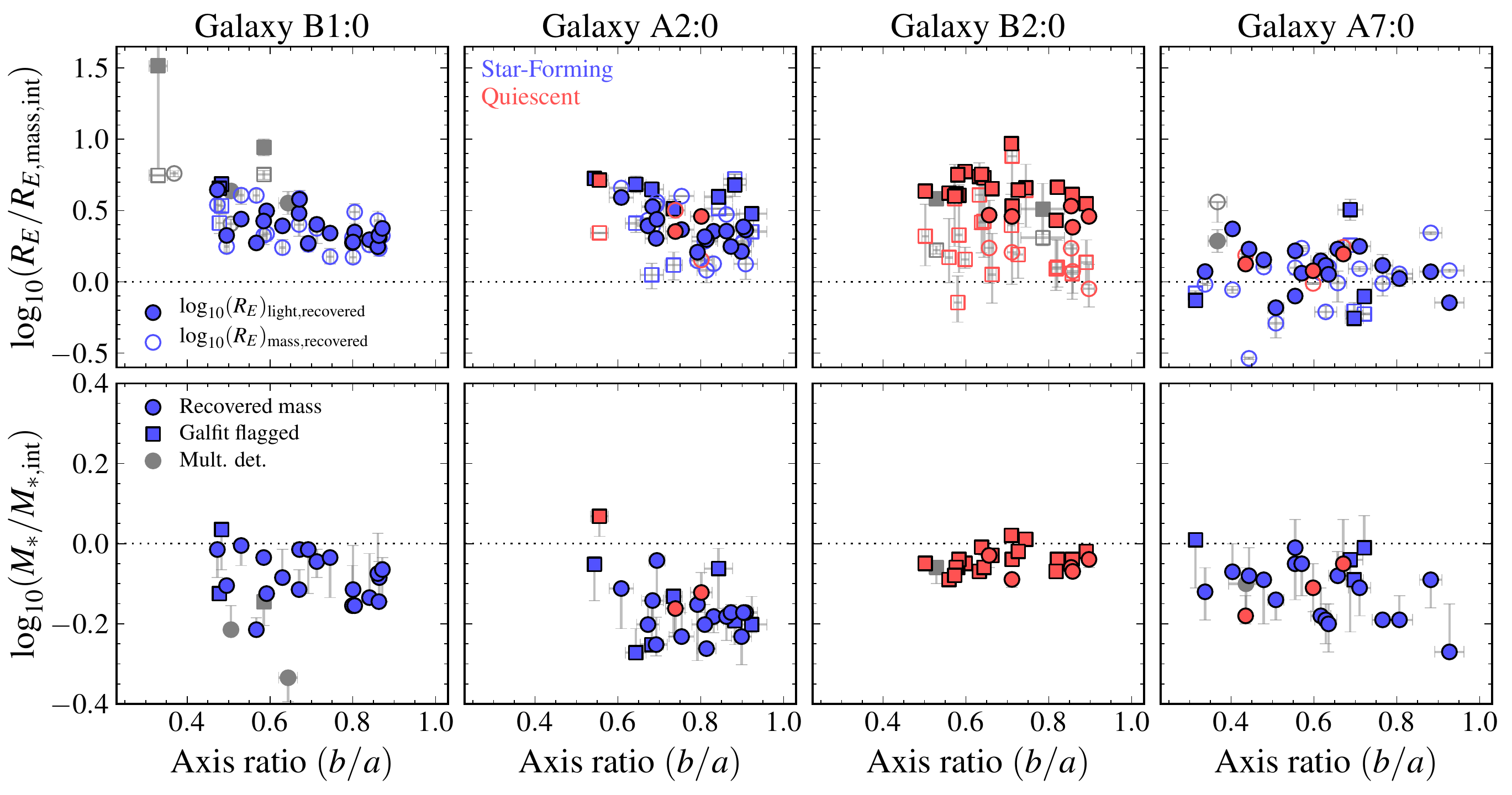}\caption{Comparison 
		of recovered and intrinsic sizes and masses 
		over 25 random viewing angles 
		for four simulated MassiveFIRE 
		galaxy snapshots with $\Mstar\sim10^{10}-10^{11}\Msun$ at $z\sim2$
		as a function of \textsc{Galfit} axis ratio.
		In the top row, we show the ratio of the 
		\textsc{Galfit} half-light radii from the rest-frame V band (filled symbols) 
		and half-mass radii following \citet{Szomoru13} (open symbols) to the intrinsic half-mass radii. 
		In the bottom row, we show the ratio of the recovered and intrinsic masses. 
		Star-forming and quiescent projections (see Section~\ref{sec:results}) 
		are colored blue and red, respectively.
		\textsc{Galfit}-flagged detections are marked with squares. 
		Projections with multiple detections are colored gray, 
		and only the largest radius or mass component is shown.
		Most orientations for Galaxy B2:0 are flagged due to 
		the S\'ersic index reaching the upper limit ($n=8$).}\label{fig:m_r_subset}\end{figure*}

\section{Recovering Sizes and Masses}
\label{sec:methods_recovery}

We measure masses and sizes from the mock images following established 
observational techniques. 
Stellar masses \Mstar are determined by 
fitting the \citet{bc03} SPS models to all bands of 
the mock photometry 
of every object using FAST \citep{Kriek09}. 
We assume a \citet{Chabrier03} IMF, a \citet{Calzetti00} dust attenuation curve, 
a delayed exponentially declining star formation history,  
and solar metallicity. 

Structural parameters of the simulated galaxies, including the effective radius $R_E$, 
S\'ersic index $n$ \citep{Sersic68}, and axis ratio $b/a$, are measured
using \textsc{Galfit} \citep{Peng10a} on the rest-frame V-band images. 
We use the \textsc{Galfit} parameter limits of \citet{vanderWel12} 
and flag and exclude from analysis projections 
for which (a) the \textsc{Galfit} and V-band total magnitudes differ by $>0.5\unit{mag}$ 
and (b) fit parameter(s) reached the enforced limit(s).
We adopt the semimajor axis $R_E$ as the half-light radius. 
We also estimate half-mass radii following \citet{Szomoru13}. 
This method uses rest-frame u- and g-band \textsc{Galfit} profiles and residuals together with 
an empirical mass-to-light ratio versus color relation to derive a stellar mass profile out to $100\unit{kpc}$.

To determine the fiducial intrinsic masses and sizes of the simulated galaxies, we measure 
the stellar masses and half-mass radii directly from the mass maps of each galaxy. 
We define the intrinsic stellar mass for each projection of each galaxy as the mass\footnote{\citet{bc03} model 
masses are used to avoid discrepancies between the recovered and intrinsic masses
due to variations in mass-loss prescriptions between 
the SPS models and the FIRE feedback model \citep{Hopkins14}, 
as testing mass-loss variations is beyond the scope of this Letter. 
These masses are calculated as the current \citet{bc03} model stellar 
mass given every star particle's age, initial mass, and metallicity.\label{fn:masses}} 
enclosed within the 
Source Extractor Kron ellipse \citep{Kron80}, masking neighboring detections. 
Thus, the recovered and intrinsic masses are defined for the same aperture \citep{Skelton14}.
The 2D intrinsic major-axis half-mass radii are defined from growth curves on the projected mass maps, 
using self-similar ellipses out to the elliptical Kron aperture for each projection as well.
We take the median over all projections to obtain the fiducial intrinsic stellar mass and 
half-mass radius for each galaxy. 
These intrinsic masses are similar to those derived by \citet{Feldmann17}, 
which are measured within a sphere of radius $0.1r_{\mathrm{halo}}$, 
but the adopted definition allows comparable aperture corrections to be measured 
from the noise-free light images and recovered mass profiles (see Section~\ref{sec:results}).

The recovered sizes and masses for four simulated galaxies 
over 25 random projections are shown in Figure~\ref{fig:m_r_subset}.
The top panel demonstrates that the measured half-light and half-mass radii are generally larger than the 
intrinsic radii, while the bottom panel shows that the recovered stellar masses are similar to the intrinsic masses. 
We observe scatter in both the recovered sizes and masses between different viewing angles. 
There is a slight trend of increasing radii with decreasing axis ratio $b/a$ for 
some galaxies, which could be caused by inclination-dependent color gradients.
Investigating this trend is beyond the scope of this \mbox{Letter.}

\section{Size and mass comparisons}
\label{sec:results}

To understand how well observations recover the sizes and masses of galaxies, 
we examine the median offset between the recovered and intrinsic sizes and masses 
for the sample of 50 MassiveFIRE galaxy snapshots, each with 25 projections. 
Furthermore, we examine whether these offsets differ for star-forming and quiescent galaxies. 
We use the empirical UVJ criterion by \citet{Muzzin13} at $z>1$ to 
classify each projection of all galaxies as star-forming or quiescent. 

\begin{figure*}\centering\includegraphics[width=0.92\textwidth,trim=0 8pt 0 4pt,clip=True]{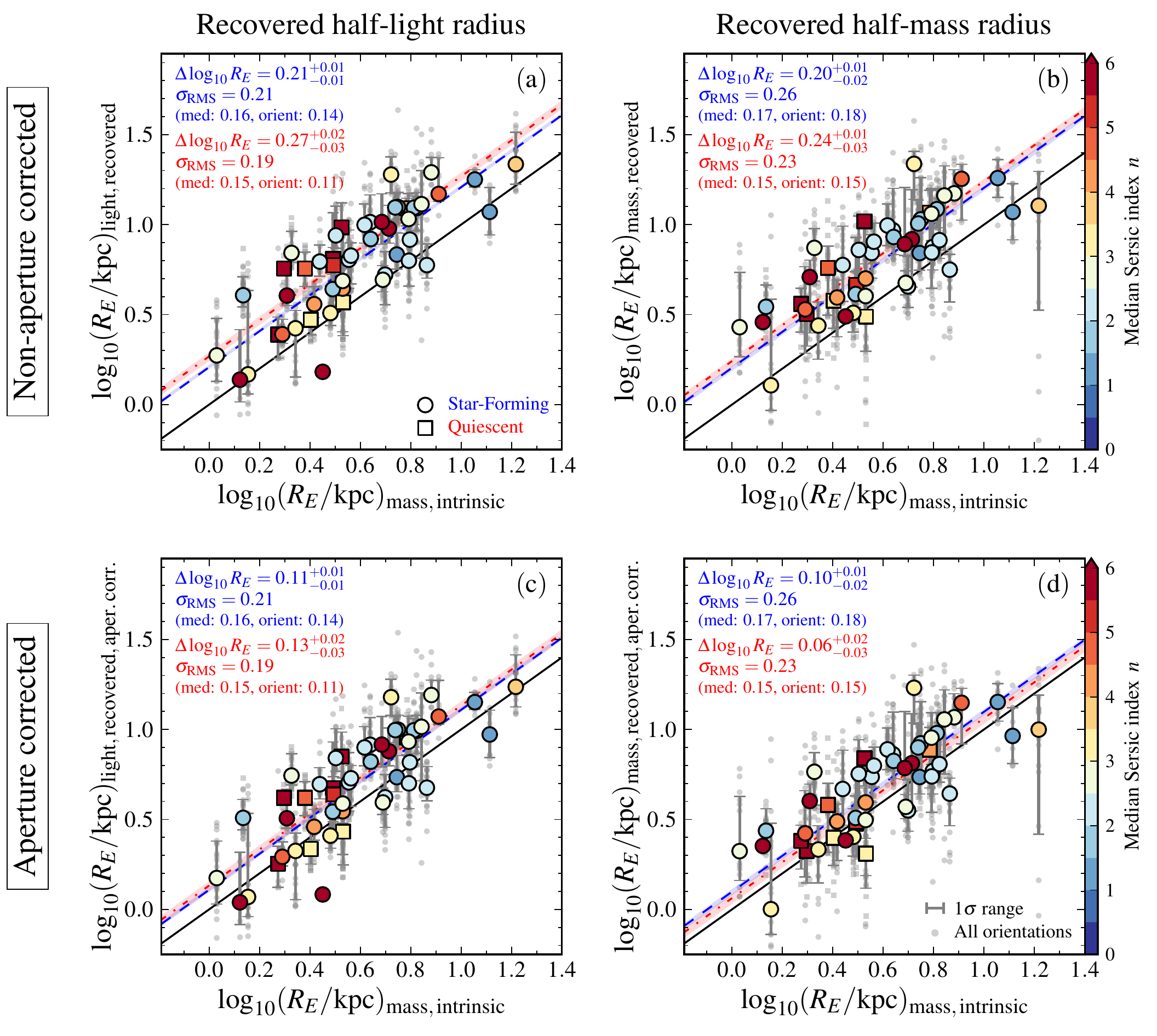}\caption{Comparison 
	between intrinsic half-mass radii and the median recovered
	(a) half-light 
	and 
	(b) half-mass radii, 
	colored by the median (across all orientations) S\'ersic index, 
	not accounting for aperture differences.
	We also compare the intrinsic half-mass radii with  
	aperture-corrected recovered 
	(c) half-light  
	and 
	(d) half-mass radii.
	Star-forming and quiescent galaxies 	are shown with circles and squares, respectively. 
	Median U-V and V-J colors are used to categorize each galaxy.
	\textsc{Galfit}-flagged detections are excluded from the median and scatter calculations.
	The black line shows the one-to-one relationship, and the star-forming and quiescent 
	median size offsets (over all projections) are shown 
	with dashed blue and dashed-dotted red lines, respectively. 
	The shaded regions show the $1\sigma$ offset uncertainties. 
	The small 	gray circles (squares) show the radii of 
	all non-flagged star-forming (quiescent) orientations, 
	and the bars denote the $1\sigma$ range of radii for each galaxy.}\label{fig:size_compare}\end{figure*}

In Figure~\ref{fig:size_compare}, we show the median recovered half-light and half-mass radii versus intrinsic radii, 
excluding all \textsc{Galfit}-flagged detections. We also show all projections and their $1\sigma$ scatter.  
We use all unflagged orientations of all galaxies to determine the median offset between the recovered 
and intrinsic sizes. The offset uncertainties are estimated by bootstrapping the error on the median.

We find that \textsc{Galfit} radii overestimate the intrinsic radii (Figure~\ref{fig:size_compare}a), 
with median offsets of $\Delta\log_{10}R_E=0.21$ and $0.27\unit{dex}$ for the 
star-forming and quiescent samples, respectively.  
The scatter in $\log_{10}R_{E,\mathrm{light,recovered}}$ 
over all projections for star-forming and quiescent galaxies 
is $\sigma_{\mathrm{RMS}}=0.21\unit{dex}$ and $0.19\unit{dex}$, respectively. 
The scatter of the median sizes of star-forming and quiescent galaxies 
($\langle\log_{10}R_{E,\mathrm{light,recovered}}\rangle$; 
weighted by the fraction of unflagged projections) is 
$\sigma_{\mathrm{med}}=0.16$ and $0.15\unit{dex}$, respectively. 
Thus, orientation increases the total scatter by 
$\sigma_{\mathrm{orient}}\sim\!0.14$ and $0.11\unit{dex}$ 
for star-forming and quiescent galaxies, respectively (accounting for measurement errors).

The recovered half-mass radii 
are also offset from the intrinsic radii (Figure~\ref{fig:size_compare}b), 
by $\Delta\log_{10}R_E=0.20$ and $0.24\unit{dex}$ 
for the star-forming and quiescent galaxies, respectively. 
The scatter of $\log_{10}R_{E,\mathrm{mass,recovered}}$ over all projections 
is slightly larger ($\sigma_{\mathrm{RMS}}=0.26$ and $0.23\unit{dex}$), 
with a similar fraction caused by orientation ($0.18$ and $0.15\unit{dex}$). 
In comparison to \citet{Szomoru13}, our sample has relatively flat u-g profiles, 
resulting in similar half-light and half-mass radii.

However, these size comparisons do not account for aperture effects.
The intrinsic half-mass radii are defined within finite elliptical apertures
(Section~\ref{sec:methods_recovery}), whereas \textsc{Galfit} S\'ersic profiles are parametric 
and integrated out to infinity. 
To quantify the aperture effects on the measured light-mass size offsets, 
we compare \textsc{Galfit} effective radii to median aperture half-light radii 
and the recovered half-mass radii to median recovered aperture half-mass radii. 
Aperture half-light radii are measured directly from 
the noise-free V-band images, analogous to the half-mass radii measurements. 
Similarly, recovered aperture half-mass radii are derived from the measured mass profiles. 
In all cases, aperture effects account for $\sim0.1\unit{dex}$ of the size offsets.

We find that aperture-corrected half-light radii are in fairly good agreement with the 
intrinsic half-mass radii (Figure~\ref{fig:size_compare}c), with larger half-light radii by
$0.11\unit{dex}$ and $0.13\unit{dex}$ for star-forming and quiescent galaxies, respectively, 
in agreement with previous studies (\citealt{Wuyts10}, \citealt{Wuyts12}, \citealt{Szomoru13}). 
The aperture-corrected half-mass and intrinsic half-mass radii have similar systematic offsets 
($0.10\unit{dex}$ and $0.06\unit{dex}$; Figure~\ref{fig:size_compare}d).

The difference between the aperture-corrected half-light and half-mass radii 
(Figure~\ref{fig:size_compare}c) appears to be caused by the presence of dust-obscured 
high central mass concentrations (within $\lesssim\!1\unit{kpc}$) in many of the galaxies. 
We would expect that observed color gradients would enable us to recover 
the central mass component. 
However, the high central dust content results in saturated color profiles, so this mass component is 
not recovered using the method of \citet{Szomoru13} (Figure~\ref{fig:size_compare}d). 
Another potential source of bias between the intrinsic and recovered radii is the use of smooth, 
single-S\'ersic models, as these galaxies have complex structures. Nonetheless, we find that 
single-S\'ersic models introduce little to no bias to the recovered sizes, 
in agreement with other studies (e.g., \citealt{Davari14, Davari16}). 

Recovered and intrinsic stellar masses are compared in Figure~\ref{fig:mass_compare}, 
using the same the median calculation method and set of non-flagged detections 
as for the size comparison. We find that the recovered masses are generally in good agreement 
with the intrinsic masses, with an offset of only $-0.06\unit{dex}$ 
for both star-forming and quiescent galaxies, and have a scatter of 
$\sigma_{\mathrm{RMS}}=0.14$ and $0.11\unit{dex}$ over all projections, 
with $0.10$ and $0.05\unit{dex}$ due to orientation effects. 
Uncertainties in stellar masses can arise from both measured photometry 
and from mass-to-light ratios derived from SED fitting. 
We find that photometric uncertainties do not strongly affect the 
accuracy of the recovered stellar masses. 
The Source Extractor-derived fluxes recover the intrinsic aperture fluxes very well, 
with a median fractional flux difference of $-0.3\%$ and an rms scatter of $7.5\%$.
The small offset and scatter show that stellar masses are recovered 
well on average over a wide mass range ($\sim10^{9.75}-10^{11.25}\Msun$)
and dust attenuation range ($A_{\mathrm{V}}\sim0-2$),
but do vary with galaxy viewing direction. 
Our result of no large systematic mass offset is in good agreement with the findings 
of other tests of stellar mass recovery using mock observations of simulations 
(e.g., \citealt{Wuyts09b}, \citealt{Torrey15}).

\section{Discussion and Implications}
\label{sec:discussion}

Using mock multi-band images of MassiveFIRE simulated galaxies, 
we show that recovered half-light radii are in good agreement with 
the intrinsic half-mass radii, with an offset of 
$\log_{10}R_{E,\mathrm{light,recovered}}-\log_{10}R_{E,\mathrm{mass,intrinsic}}\sim\!0.1\unit{dex}$ 
(correcting for aperture effects). 
When we recover half-mass radii by accounting for 
color gradients due to dust, metallicity, and age, 
the radii have a similar offset of \hbox{$\sim\!0.1\unit{dex}$}.
Stellar masses are also recovered well on average, with an offset of 
$\log_{10}M_{*,\mathrm{recovered}}-\log_{10}M_{*,\mathrm{intrinsic}}\sim\!-0.06\unit{dex}$.

By considering the multiple viewing angles of every galaxy, we show that a 
sizable fraction of the mass and radii scatter is caused by orientation effects.
These projection effects may result from the random distribution of 
bright clumps within a galaxy, a non-uniform or patchy dust distribution, 
or gradients in metallicity and stellar population age \citep{Kelvin12}.

We find no systematic differences between the recovery of 
masses or radii for massive star-forming and quiescent galaxies. 
Thus, observed differences between star-forming and quiescent galaxy 
sizes at $z\sim2$ likely indicate true differences in their stellar mass distributions.

\begin{figure}\centering\includegraphics[width=0.45\textwidth,trim=0 10pt 0 2pt,clip=True]{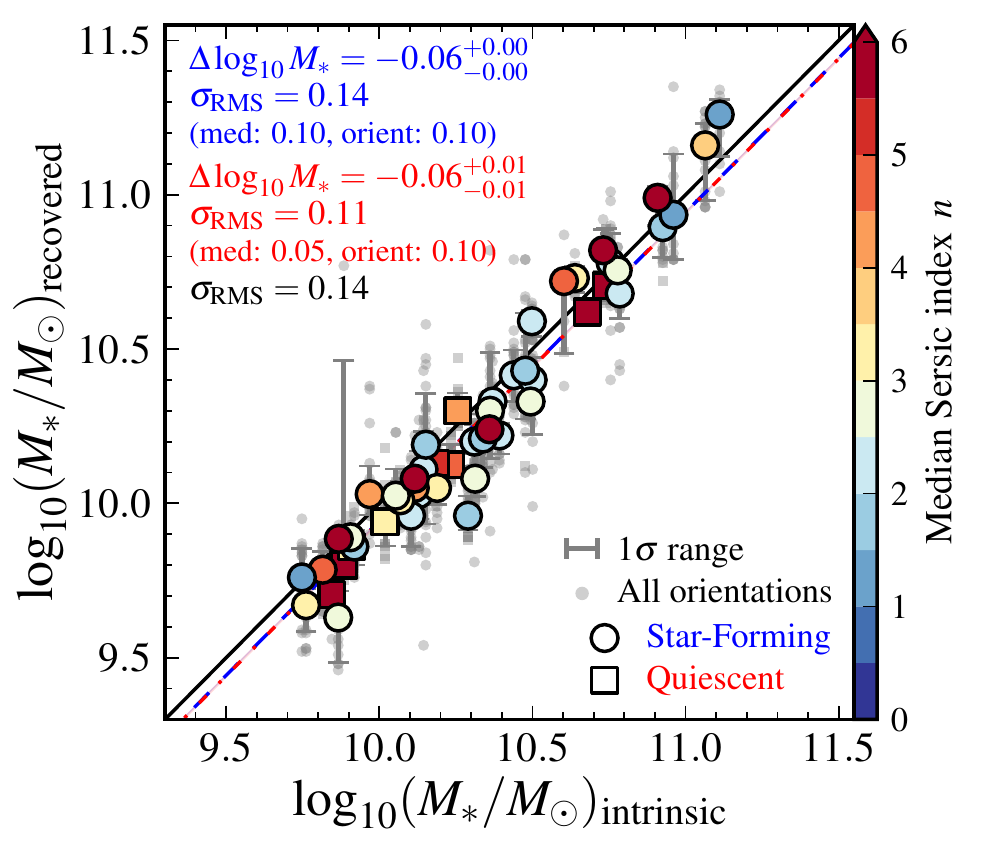}\caption{Comparison 
	between intrinsic and median recovered stellar masses, colored by the median S\'ersic index. 	
	The symbol definitions are the same as in Figure~\ref{fig:size_compare}, 
	and again \textsc{Galfit}-flagged detections are excluded.  
	The median offset between the recovered and intrinsic masses is small 
	for both star-forming and quiescent galaxies.}\label{fig:mass_compare}\end{figure}

\begin{figure*}[!t]\centering\includegraphics[width=0.825\textwidth,trim=0 8pt 0 0,clip=True]{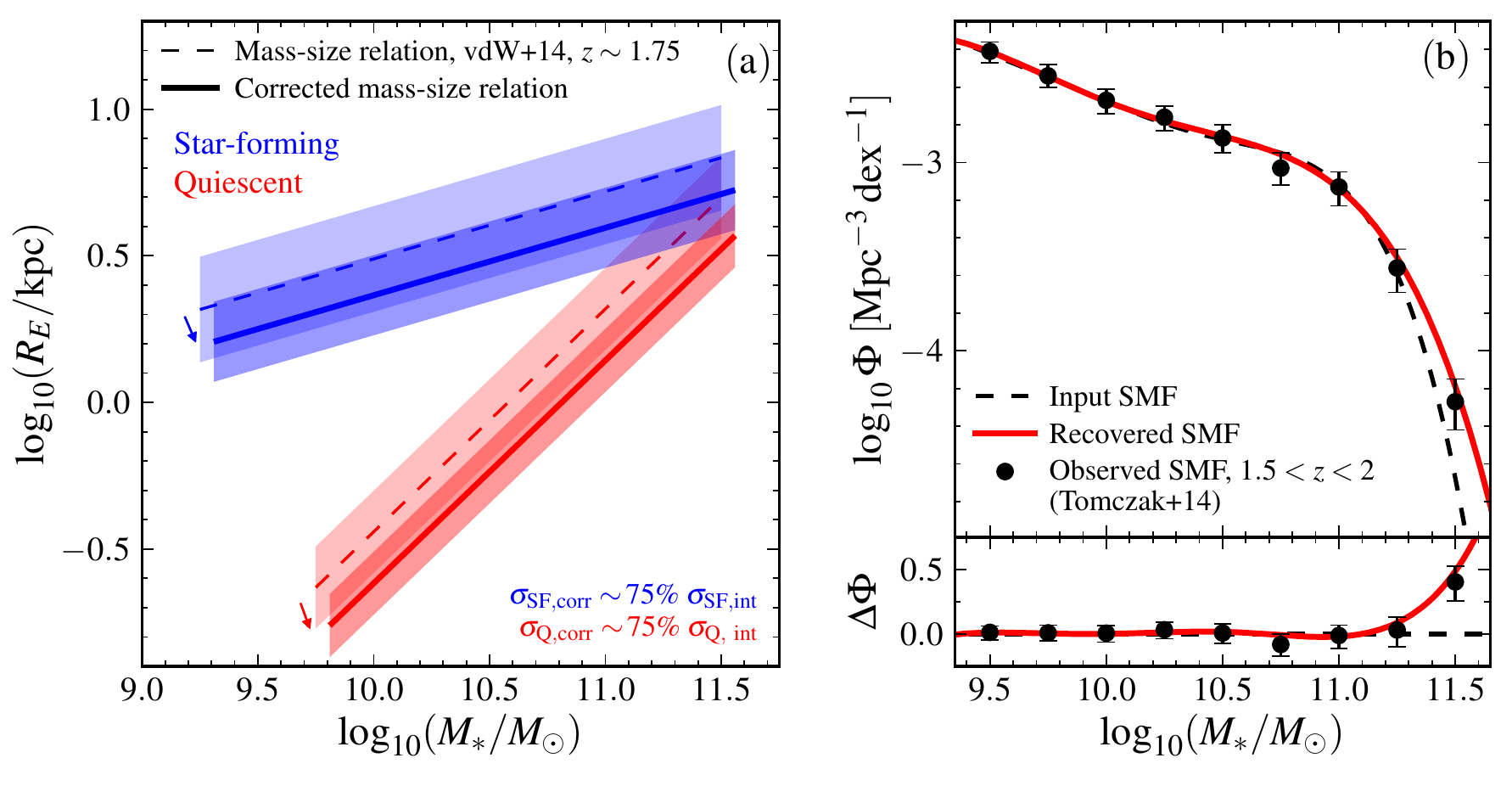}\caption{Possible 
	systematic effects on observed galaxy properties at $z\sim1.7-2$. 
	 (a) Shift and scatter reduction of observed mass-size relations.
	 Dashed lines show the star-forming (blue) and quiescent (red) mass-size relations by 
	 \citet{vanderWel14a} at $z\sim1.75$, 
	 with light shaded regions showing the observed intrinsic scatter $\sigma\log_{10}(R_{\mathrm{eff}})$. 
	 Corrected mass-size relations (based on half-mass radii) are shown with solid lines, 
	 demonstrating the inferred zero-point offset (Figures~\ref{fig:size_compare}~and~\ref{fig:mass_compare}). 
	 The darker shaded regions represent the intrinsic scatter corrected for orientation effects. 
	 (b) Effect of orientation-based scatter on stellar-mass functions (SMFs). 
	The recovered SMF (red line), similar to the SMF at $1.5<z<2$ observed by \citet{Tomczak14} (black circles), 
	 deviates from the true SMF (black dashed line) 
	 at high masses when mass scatter is applied to 
	 a sample drawn from the true SMF.}\label{fig:mass_size_shift}\end{figure*}

These results have important implications for measuring galaxy structural growth through 
mass-size relations. First, the mass-size relation zero-point will be systematically 
overestimated by $\sim\!0.1\unit{dex}$ if half-light radii are used rather than half-mass radii. 
Second, the intrinsic scatter of the light-based mass-size relation 
may be overestimated due to random variations in viewing angle, 
implying the intrinsic mass-size relation could be tighter than previously thought. 
To quantify the effect of orientation on the mass-size relation scatter, 
we compare the combined orientation-corrected mass and radius scatters
with the combined total scatter. 
We use the scatter of the medians, $\sigma_{\mathrm{med}}$, 
as the ``intrinsic'' scatter (as the mass and radii offsets are uncorrelated), 
and take the error-corrected RMS scatter as the total scatter, 
$\sigma_{\mathrm{tot}}=\sqrt{\sigma_{\mathrm{RMS}}^2-\sigma_{\mathrm{err}}^2}$. 
The orientation-corrected mass-size relation scatter is $\sim\!75\%$ 
of the error-corrected total scatter ($\sigma_{\mathrm{med}}/\sigma_{\mathrm{tot}}$) 
for both star-forming and quiescent galaxies.

We illustrate the differences between the observed mass-size relations at $z\sim1.75$ 
by \citet{vanderWel14a}
and the inferred half-mass radii mass-size relations corrected for orientation effects
for both star-forming and quiescent galaxies
in Figure~\ref{fig:mass_size_shift}a. 
This figure demonstrates both  the zero-point offset 
due to using intrinsic half-mass versus recovered half-light radii 
(corrected for aperture effects; Figure~\ref{fig:size_compare}) 
and stellar mass recovery (Figure~\ref{fig:mass_compare}), 
and the reduced intrinsic scatter once orientation effects are corrected.

Even though masses are recovered well on average, the scatter in stellar masses has 
important implications for studying galaxy populations. 
For example, scatter impacts the measurement of stellar-mass functions (SMFs). 
In Figure~\ref{fig:mass_size_shift}b, we demonstrate how orientation scatter causes an
overestimate of the number density of high-mass galaxies. 
We draw a galaxy population directly from an input SMF, 
perturb the masses by the orientation scatter, and then measure the SMF. 
The input parameters are chosen so the recovered SMF roughly 
approximates the best-fit $1.5<z<2$ SMF by \citet{Tomczak14}.
The true SMF falls off faster than the observed SMF at high masses due to 
the combination of projection-driven scatter and the steepness of the SMF at the high-mass end, 
by up to \hbox{$\sim\!0.5\unit{dex}$ at $M_*\sim10^{11.5}\Msun$}. 
Hence, many massive galaxies may have such large observed masses 
as a result of orientation effects.
Orientation-driven scatter will also impact other measurements, 
including the scatter of the star-forming main sequence 
(e.g., \citealt{Whitaker14a}, \citealt{Shivaei15}) and inferred 
dynamical masses (e.g., \citealt{Price16}, \citealt{Wuyts16}).

Furthermore, our results demonstrate the difficulty of comparing 
the sizes of observed and simulated galaxies (see Figure~\ref{fig:size_compare}a).
When directly comparing 3D-aperture half-mass radii derived from 
the simulations and \textsc{Galfit} effective radii, we find an offset of $\sim\!0.2\unit{dex}$ 
for both star-forming and quiescent galaxies. 
To make a fair comparison between observations and simulations, 
simulated galaxy half-light radii should be measured from mock images 
using the same methodology applied to observations.

We note the following caveats to this analysis. 
First, the selected galaxies may not be fully 
representative of the properties of massive galaxies at \hbox{$z\sim1.7-2$}.
Thus, the measured offsets may not be applicable to all galaxies at these redshifts. 
Moreover, the relative corrections for star-forming and quiescent galaxies 
may depend on the realism of the specific simulation models. 
Finally, we do not account for systematic modeling errors. We have only considered one 
set of stellar population models and one dust law, applied with a simple line-of-sight attenuation. 
Modeling choices could affect the recovered offsets and the scatter through systematic 
color gradient trends and variation in dust attenuation over different viewing angles.
Future work is needed to fully understand the impact of dust, non-smooth galaxy morphologies, and 
specifics of the dust radiative transfer modeling when measuring simulated galaxy properties.



\acknowledgements
We acknowledge valuable discussions with 
M.~Franx, D.~Szomoru, K.~Whitaker, C.~Hayward, C.-P.~Ma, and K.~Suess.
This work made use of \texttt{astropy} \citep{Astropy13} and \texttt{pysynphot} \citep{Lim15}. 
S.P was supported by a National Science Foundation Graduate Research Fellowship under grant DGE 1106400.  
M.K. acknowledges support from NSF AAG grant 1313171 and STScI grants AR-13907 and AR-12847, 
provided by NASA through a grant from the Space Telescope Science Institute. 
R.F. was supported in part by NASA through Hubble Fellowship grant HF2-51304.001-A awarded by 
the Space Telescope Science Institute, which is operated by the 
Association of Universities for Research in Astronomy, Inc., 
for NASA, under contract NAS 5-26555, by the 
Theoretical Astrophysics Center at UC Berkeley, 
and by the Swiss National Science Foundation (grant No. 157591). 
R.F. and E.Q. acknowledge support from NASA ATP grant 12-ATP-120183. 
P.H. was supported by an Alfred P. Sloan Research Fellowship, NASA ATP grant NNX14AH35G, 
and NSF Collaborative Research grant 1411920 and CAREER grant 1455342. 
C.A.F.G. was supported by NSF grants AST-1412836 and AST-1517491, 
NASA grant NNX15AB22G, and STScI grant HST-AR-14562.001. 
D.K. acknowledges support from NSF grant AST-1412153 and a Cottrell Scholar Award from the RCSA.

\bibliographystyle{apj}

\end{document}